\documentclass[twocolumn]{aastex62}

\newcommand{\jiang}{\color{black}}
\newcommand{\xie}{\color{black}}
\newcommand{\xx}{\color{black}}

\newcommand{\reffig}[1]{Figure \ref{#1}}

\graphicspath{{./}{figures/}}

\shorttitle{Original article}
\shortauthors{Jiang et al.}

\begin{document}

\title{On the Orbital Spacing Pattern of Kepler Multiple Planet Systems}

\correspondingauthor{Ji-Wei Xie}
\email{jwxie@nju.edu.cn}

\author{Chao-Feng Jiang, Ji-Wei Xie and Ji-Lin, Zhou}
\affiliation{School of Astronomy and Space Science \& Key Laboratory of Modern Astronomy and Astrophysics in Ministry of Education, Nanjing University, 210093, China}

\begin{abstract}
The {\it Kepler} space mission has detected a large number of exoplanets in multiple transiting {\jiang planet} systems.
Previous studies found that these Kepler multiple planet systems exhibit an intra-system uniformity, namely planets in the same system have similar sizes and correlated orbital spacings.
{\jiang However, it is important to consider the possible role of selection effects due to observational biases.}
In this paper, {\xie we revisit the orbital spacing aspect of the pattern  after taking observational biases into account using a forward modeling method}. 
We find that orbital spacings, in terms of period ratios, of Kepler multiple planet systems are significantly correlated only for those tightly packed systems, and the transition from correlation to non-correlation is abrupt with a boundary at mean period ratio $\overline{PR}$ $\sim 1.5 - 1.7$.
In this regard, the pattern of orbital spacing is more like a dichotomy rather than a global correlation. 
Furthermore, we find that such an apparent orbital spacing dichotomy could be essentially a projection of a dichotomy that related to mean motion resonance (MMR), which we dub as MMR dichotomy, and itself could be a natural result of planet migration and dynamical evolution.
%{\jcf In addition, we find that the dichotomy pattern is attributed to neither missing planets nor ultra-short period planets. 
%Instead, by reproducing observed results using a two population model, we further demonstrate the dichotomy of the orbital spacing pattern.}

\end{abstract}


\section{Introduction}  \label{intro}
Hitherto, the number of detected exoplanets has been boosted to over 4000 thanks to various ground-based and space-based surveys, among which the {\it Kepler} mission \citep{Bor10} has played a major role in contributing over two thirds of these discoveries\footnote{http://exoplanet.eu}.
The bulk of exoplanets detected by the {\it Kepler} mission are so called super-Earth or sub-Neptunes with radii between Earth and Neptune and orbital periods less than several hundred days \citep{Tho18}.
Although super-Earths are found to be common \citep{DZ13,How13,Zhu18,Mul15}, they do not exist in our Solar System, and how they were formed remains an open question.\citep{Lis14,MR16}

Among the Kepler discoveries, one of the most valuable parts is the large sample of multiple transiting planet systems \citep{RH10}, which has greatly advanced our knowledge on exoplanets in many aspects, including planetary masses and thus physical compositions \citep{Car12,HL14,WL13}, orbital eccentricities and inclinations \citep{FM12,Fab14,Xie16,Van19} and etc., shedding light on their formation and evolution history \citep{Mil16,OC20}.

In this paper, we focus on the aspect of orbital spacing, which has attracted numerous studies. 
 \citet{BL13} and \citet{Hua14} investigated the orbital spacings of Kepler's multiple systems in a context of extended Titus-Bode law of our Solar System.
\citet{PW15} found that the orbital spacings of Kepler planets are clustered around the theoretical stability threshold. 
Some studies investigated the spacings of Kepler planets in terms of orbital period ratio \citep{Lis11,Ste13,SH15}.
From the period ratio distribution, the majority of Kepler planets were found to be not in mean motion resonance (MMR).
{\jiang Nevertheless, the period ratio distribution has shown overabundances just wide of first-order MMRs and deficits short of them \citep{Fab14}, which may have implications to planet formation and evolution \citep{LW12,BM13,Xie14,DL14,CF15,ML19}. }

Recently, \citet{Wei18} found that planets orbiting the same host tend to be similar in sizes (see also in \citet{Mil17,Wang17} ) and have regular orbital spacings (i.e., period ratio correlation), a pattern which they dubbed as `peas in a pod'.
However, whether such a pattern is astrophysical or a selection effect due to observational biases is still currently in debate \citep{Zhu20,WP20,MT20,GF20}. %{\xie (update Zhu 2019 and WP19 to Zhu 2020, and WP 2020, and add two more refs here, i.e., Murchikova, L., & Tremaine, S. 2020 arXiv:2003.02290 and https://arxiv.org/abs/2003.11098)}

Here, we revisit {\xie one aspect of the pattern, i.e.,} the period ratio correlation, in detail by taking observational biases into account. 
This paper is organized as follows.
In section \ref{sample}, we select different planet samples by applying different criteria. 
Then, for each planet sample, we evaluate the significance of period ratio correlation and the effects of observational biases (section \ref{result.rev_prcor}).
We find evidences, in section \ref{result.evi_prdich}, which show that the orbital spacing pattern is more like a dichotomy rather than a global correlation.
In section \ref{discuss}, we discuss the implications of such an orbital spacing dichotomy. 
Section \ref{summary} is the summary of the paper.

\section{Sample} \label{sample}
Our study is based on the multiple transiting planet systems detected by the {\it Kepler} mission. 
We use the Q1-Q17 table of Kepler Objects of Interest (hereafter KOIs) from the NASA Exoplanet Archive.   \footnote{https://exoplanetarchive.ipac.caltech.edu}. 
Firstly, we exclude all the KOIs which are identified as false positives.
Secondly, we adopt three filters as follows to the remaining planetary systems. 
\begin{itemize}
    \item[1] The multiplicity of planetary systems $N_p \geqslant 4$.
    \item[2] The maximum radius of planets in the systems $R_{max} \leqslant 6R_{\oplus}$, where $R_{\oplus}$ is the Earth radius
    \item[3] The maximum of period ratios of adjacent planets in the systems $PR_{max} \leqslant 4.0$
\end{itemize}
{\xie We adopt the first filter for the reason that systems of lower multiplicities tend to be not dynamically packed and thus have a higher likelihood of missing non-transiting planets in between the transiting planets (see more discussions in section 4.3.1), causing a systematic overestimation of the period ratios of neighbouring planets.}
Through the second filter, we exclude giant planets, allowing us to focus on smaller planets, i.e. super-Earths and sub-Neptunes.
We adopt the third filter according to \citet{Wei18} for comparison with their results.
After all these three filters, we have 56 multiple planet systems in our nominal sample (sample 1 in table \ref{Tab.samp_des}).

{\xie For comparison with sample 1, we adjust the above three filters to construct our sample 2 and sample 3.
In sample 2, we release the radius cutoff to include those systems which host giant planets with $R_{max}>6R_{\oplus}$.
In sample 3, we release the spacing cutoff of $PR_{max} \leqslant 4.0$.
Besides, we adopt the same sample of  \citet{Wei18} as our sample 4.
The descriptions of the samples are summarized in table \ref{Tab.samp_des}.
}
% \usepackage{booktabs}
\begin{table}[]
\centering
\begin{tabular}{@{}cccccccc@{}}
\toprule
Sample id  & 1 &2 &3 & 4 (Weiss+ 2018) \\
\colrule
$R_{max} \leqslant 6 R_{\oplus}$ & Yes &No &Yes  & No\\
$PR_{max} \leqslant 4$ &Yes &Yes &No & Yes \\
$N_{sys}$ &56 &60 & 65 & 95\\
$N_{3}$ &0 &0 & 0 & 53\\
$N_{4}$ &39 &41 & 46 & 31\\
$N_{5}$ &15 &17 & 17 & 10\\
$N_{6}$ &2 &2 & 2 & 1\\
\botrule
\end{tabular}
\caption{ {\xie Summary of the samples 1, 2, 3  and 4 mentioned in section \ref{sample}. Different cut conditions (i.e., $R_{max} \leqslant 6R_{\oplus}$ and $PR_{max} \leqslant 4$) are applied to some of the samples. $N_{sys}$ is the total number of systems and $N_3,N_4,N_5$ and $N_6$ are the specific numbers of systems with 3,4,5 and 6 transiting planets respectively.}}
\label{Tab.samp_des}
\end{table}

\begin{figure}%[tbhp]
\centering
\includegraphics[width=\linewidth,height=\linewidth]{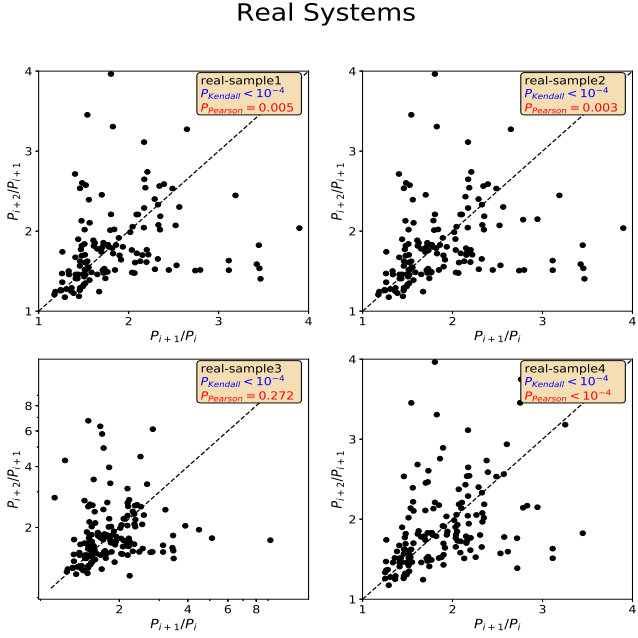}
\caption{The period ratio correlation evaluation for the four observed samples (see table \ref{Tab.samp_des} and section \ref{sample}). The x-axis and the y-axis in each panel denote the period ratio of the inner pair of neighbouring planets ($P_{i+1}/P_{i}$) and that of the outer pair ($P_{i+2}/P_{i+1}$). On the {\xie upper-right} of each panel, we printed the P value of {\jiang the Kendall correlation test and Pearson correlation test} (section \ref{result.rev_prcor.eval}). The {\xie grey dashed line} shows the perfect correlation, i.e. $y=x$. We can see that {\jiang all the samples show strong PR correlation in the Kendall correlation test. However, in the Pearson correlation test, all the sample except Sample 3 show strong PR correlation. {\xie We note that the relatively weaker PR correlation in the sample 3 is probably attributed to the inclusion of planet pairs with larger period ratios, i.e., $PR>4$} (The axes scale in the bottom-left panel is different from 
the other panels.). In fact, the trend that the period ratio correlation becomes weaker with increasing period ratio can be indeed seen in all the samples.}}
\label{obs_pr}
\end{figure}

\section{Results} \label{result}
\subsection{Revisit the Period Ratio Correlation} \label{result.rev_prcor}
First, we revisit the period ratio correlation \citep{Wei18} in different samples in section \ref{sample}.
\subsubsection{Correlation Evaluation} \label{result.rev_prcor.eval}
%briefly introduction of the results of Wei18
In the work of \citet{Wei18}, the authors measured the correlation of the orbital period ratio of each pair of neighbouring planets $P_{i+1}/P_{i}$ and that of the outer pair of neighbouring planets $P_{i+2}/P_{i+1}$.
They found a Pearson-R correlation coefficient of 0.46 with a significance of P value $<10^{-5}$, leading to a conclusion that there is a strong correlation among orbital period ratios of planets in the same systems.
{\jiang Pearson correlation coefficient, however, is not very appropriate for searching correlations in a relatively small sample, because it assumes the linear correlation and Guassian scatter.
{\xie For this reason, besides using the  Pearson correlation coefficient (mainly for comparison with \citet{Wei18}), we further repeat all the analyses using the Kendall's tau correlation coefficient, which is non-parametric without making neither assumptions, and thus more robust.}}
The detailed procedure is as follows.
\begin{itemize}
    \item[Step 1] We calculate Kendall's tau nonparametric correlation coefficient $\tau_{obs}$ {\jiang (or Pearson's correlation coefficient $R_{obs}$)} for each sample in section \ref{sample}. 
    \item[Step 2] We randomly scramble period ratios of neighbouring planets among planetary systems then re-calculate the correlation coefficient for each simulated realization $\tau_{sim}$ (or $R_{sim}$).
    \item[Step 3] We repeat Step 2 for 10000 times and calculate the fraction of times with $\tau_{sim}>=\tau_{obs}$. (or $R_{sim}>=R_{obs}$) This fraction gives the P value $P_{Kendall}$ (or $P_{Pearson}$) of the Kendall Correlation Test and $1-P_{Kendall}$ (or $1-P_{Pearson}$) is the confidence level of the observed period ratio correlation. 
\end{itemize}

\begin{figure}%[tbhp]
\centering
\includegraphics[width=\linewidth,height=\linewidth]{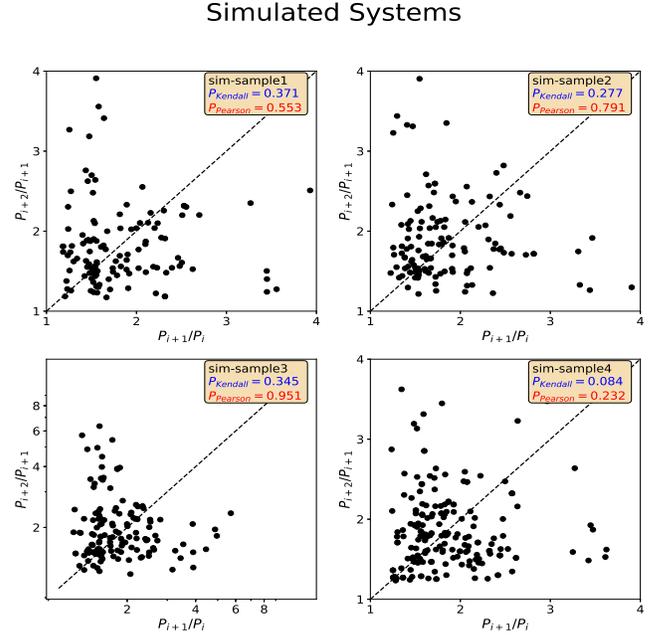}
\caption{Similar to \reffig{obs_pr}, but for a set of typical Monte Carlo realizations of simulated samples with the assumption that planets are intrinsically randomly paired (see section \ref{result.rev_prcor.eff_obsbias}).  Compared to \reffig{obs_pr}, the period ratio correlations vanish, with much larger P values of the Kendall(Pearson) correlation tests, $P_{Kendall}$($P_{Pearson}$).}
\label{typi_simu_pr}
\end{figure}
%result report Fig.1.

\begin{figure}%[tbhp]
\centering
\includegraphics[width=1\linewidth,height=1\linewidth]{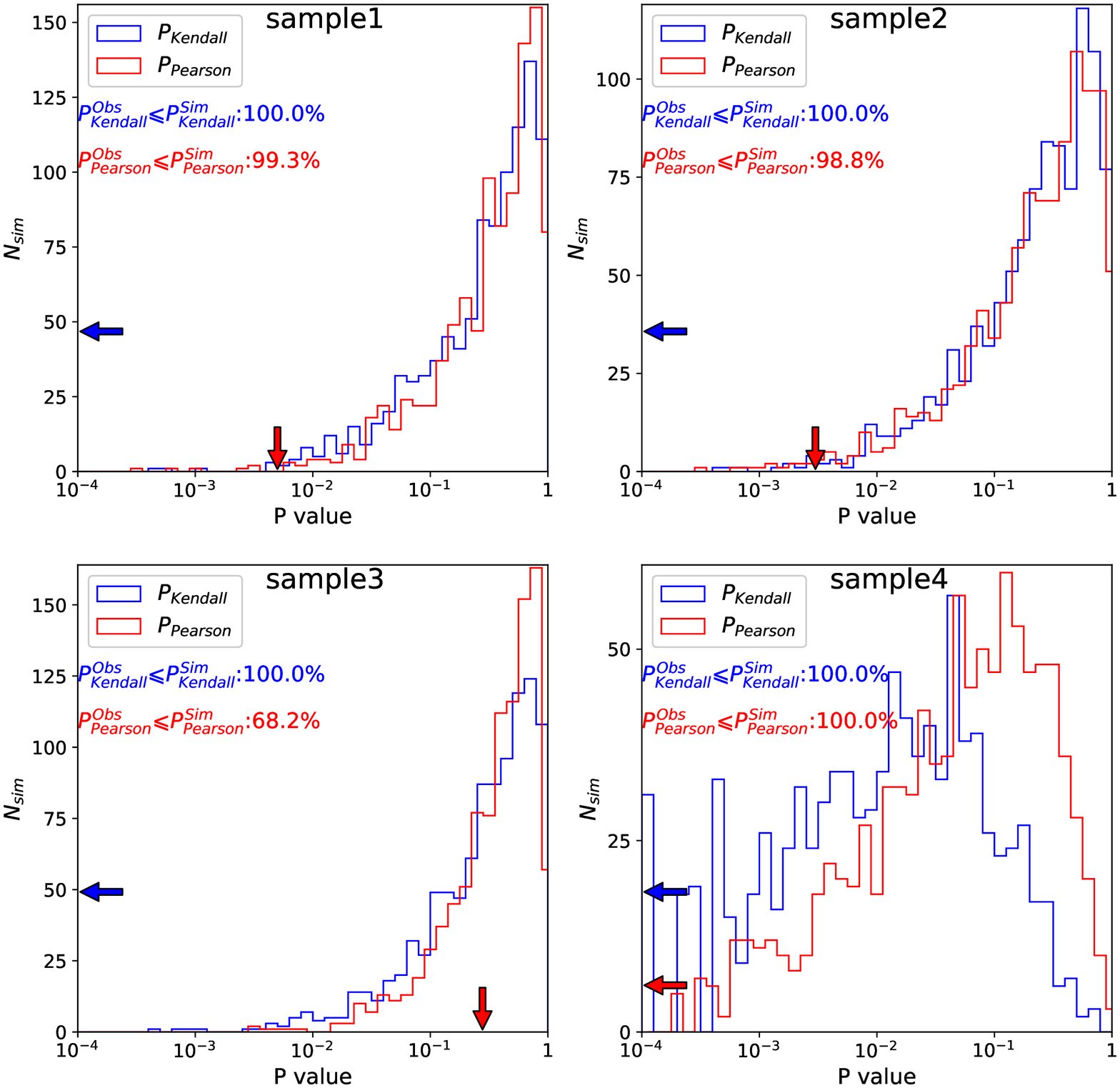}
\caption{
{\xie The distribution of $P_{Kendall}$ (blue histograms) and $P_{Pearson}$ (red histograms) for 1000 Monte Carlo realizations (see section \ref{result.rev_prcor.eff_obsbias} and the appendix) of simulated samples.  The arrows in each panel show the $P_{Kendall}$ (blue) and $P_{Pearson}$ (red) of the corresponding observed sample. 
In each panel, we print the fractions of simulations whose P values are not smaller than the observed ones, i.e., $P_{Pearson}^{obs} \leqslant P_{Pearson}^{sim}$, $P_{Kendall}^{obs} \leqslant P_{Kendall}^{sim}$, which can be treated as the confidence level that the observed correlation cannot be reproduced by observational biases.
}
}
\label{simu_pr_dis}
\end{figure}

\reffig{obs_pr} shows the period ratio correlation evaluation for the four samples as defined in table \ref{Tab.samp_des}.
For each of the sample described in table \ref{Tab.samp_des}, the period ratio correlation is significant with a confidence of larger than 99.99\% in Kendall correlation test, which is consistent with the result in \citet{Wei18} although we use different samples and correlation tests.
{\xie However, as for the bottom left panel for the result of sample 3, the Pearson test returns a much larger P value of 0.274.
This is probably because planets pairs with larger period ratios i.e., $PR>4$, are included in sample 3.
In fact, each panel also shows an apparent tend that the points with larger period ratios become more dispersed with respect to the 1:1 (y=x) line.
}
%启后
In section \ref{result.evi_prdich}, we will investigate this trend in more detail.

{\xie Note, although P values are reported to high precision here, one should not over interpret the numbers in high precision. \citep{DL11,Laz14}.  For example, $P_{Kendall}=0.279$ and $P_{Kendall}=0.378$ are essentially the same; both indicate no correlation at all. 
What really matters is the order of magnitude of the P value.}

\subsubsection{Effect of Observational Biases} \label{result.rev_prcor.eff_obsbias}

Before reaching any conclusion, one should address the issue of observational bias. 
How do the transit selection effect and detection efficiency affect the observed orbital spacing pattern? 
Could the observed pattern (\reffig{obs_pr}) be reproduced by the observational bias  \citep{Zhu20} ? 

Here, we address this issue by forward modeling the transit detection and selection process with a Monte Carlo method (see the appendix for the detailed procedure). 
With this method, we create 1000 corresponding simulated sample of equal size as each observed sample. 
We then perform the same period ratio correlation evaluation (section \ref{result.rev_prcor.eval}) to the simulated samples. 
Figure \ref{typi_simu_pr} shows the typical result of each set of simulated samples. 
As can be seen, {\jiang all the Pearson test P values for the Monte Carlo realizations are larger than 0.1, and all the Kendall test  P values are larger than 0.05}, indicating almost no correlation at all.

{\xie 
In Figure \ref{simu_pr_dis}, we plot the distributions of P values for the four simulated sample sets, and calculate the fractions of simulations whose P values are not smaller than the observed ones.
As can be seen,  in most cases (except the Pearson test in sample 3) the fraction numbers are close to $100\%$, implying high confidence level that the period ratio correlations observed in these samples are likely to be physical rather than the results of observational biases.
As for the low fraction number ($68.6\%$) for the Pearson test in sample 3, this is because the inclusion of larger period ratios largely reduces the period ratio correlation as mentioned in Figure \ref{obs_pr}.  
In the following section, we will investigate how the period ratio correlation changes with period ratio itself.
}

\subsection{Evidence of Period Ratio Dichotomy} \label{result.evi_prdich}
%对这节主要内容归纳和简介
{\xie In this subsection, we further perform a `moving sample' analysis, which reveals that the orbital spacing pattern as a whole is more like a dichotomy rather than a correlation.}
For the sake of clarity, hereafter, we only present the results of analyzing the nominal sample (sample 1 in table \ref{Tab.samp_des}), since other samples generally give similar results. 

\begin{figure*}%[tbhp]
\centering
\includegraphics[width=\linewidth,height=0.6\linewidth]{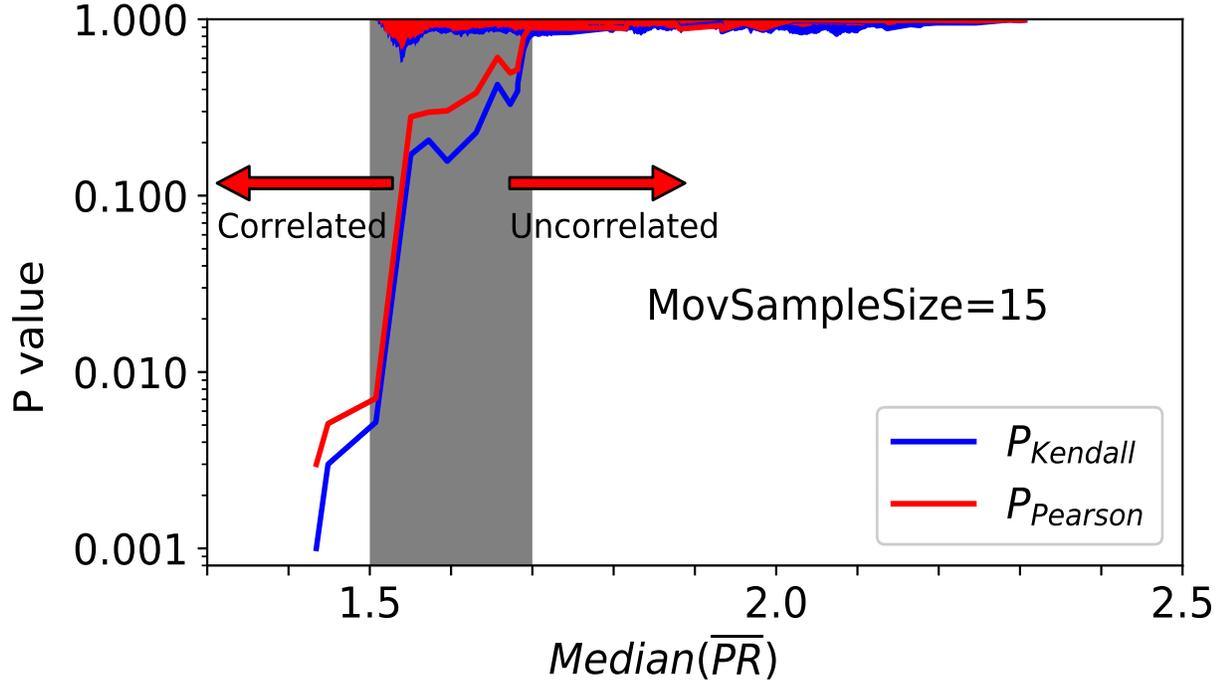}
\caption{The P value of correlation test $P_{Kendall}$ (blue solid line) and $P_{Pearson}$ (red solid line) as a function of the median value of average system period ratio of the moving subsamples, $Median(\overline{PR})$, for the nominal sample (see table \ref{Tab.samp_des}).
{\xie We see that there is an abrupt increase in the observed $P_{Kendall}$ (and $P_{Pearson}$) from $<0.01$ to $\sim 1$ at $Median(\overline{PR})\sim1.5-1.7$ (the grey shaded transition area), forming a dichotomy, namely, period ratios are correlated to each other on the left but uncorrelated on the right. 
For comparison, we also plot the 1-$\sigma$ region of the results for the corresponding simulated samples(blue and red shaded region on the top of the Figure). 
In contrast to the observed one, most simulated $P_{Kendall}$ (and $P_{Pearson}$) stay above 0.9 (i.e., uncorrelated at all) regardless of $Median(\overline{PR})$, demonstrating that the period ratio correlation, especially the dichotomous feature, could not be produced by random pairing nor by selection effects. }
}
\label{mov_sample_test}
\end{figure*}

\begin{figure*}%[tbhp]
\centering
\includegraphics[width=1\linewidth,height=0.25\linewidth]{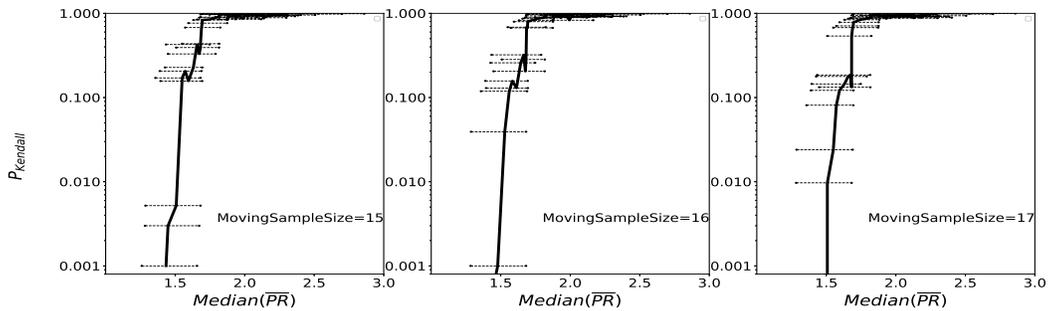}
\caption{Similar to \reffig{mov_sample_test} but here compare the results of using different sub-sample sizes (15, 16 and 17 from left to right).
For clarity, only the results for the Kendall correlation test are shown.
The dotted line across each data point shows the range of $\overline{PR}$ of individual systems in the corresponding sub-sample. All the three curves show a similar trend that $P_{Kendall}$ abruptly increases from $< 0.01$ to $\sim 1$ at $Median(\overline{PR})\sim1.5-1.7$.
{\jiang Note, in the middle and the right panel  $P_{Kendall}$ are smaller than 0.001 for $Median(\overline{PR})<1.5$, thus not shown there.}}
\label{diff_subsample_size}
\end{figure*}

The procedure of such an analysis is described as follows:
\begin{itemize}
    \item[1] Firstly, we sort all the systems in the sample according to the average period ratio of neighbouring planets $\overline{PR}$ of each system.
    
    \item[2] Secondly, we select the first 15 systems as a subsample and perform the Kendall (and Pearson) correlation evaluation (section \ref{result.rev_prcor.eval}) to the sub-sample, obtaining the P value $P_{Kendall}$ and $P_{Pearson}$.
    
    \item[3] Thirdly, we repeat the above  correlation evaluation to a series of continuously moving subsamples until the entire sample goes through. Specifically, for each time, we move the subsample one step towards larger $\overline{PR}$. For example, we select 15 systems from the 2nd and the 16th in the sorted sample next time. 
\end{itemize}

%results of the observed sample
In \reffig{mov_sample_test}, we plot the result of the above moving sample analysis, which is the P value of correlation test $P_{Kendall}$ and $P_{Pearson}$ as a function of the median of $\overline{PR}$ in each moving subsample. 
As can be seen, the P value ($P_{Kendall}$ (blue solid curve) and $P_{Pearson}$ (red solid curve)) increases from $\sim10^{-3}$ (strong correlation) to $\sim1$ (no correlation at all) as the subsample moves towards larger period ratios. 
{\xie However, the increase in P value is not smooth. The transition from correlated to uncorrelated is abrupt. 
The P value increases by more than two orders of magnitude (from $\sim0.005$ to $\sim0.8$) as the median $\overline{PR}$ just slightly changes from  $1.5$ to $1.7$. 
This transition zone (the grey shaded area in \reffig{mov_sample_test}) separates two populations; one with correlated period ratios and the other with uncorrelated ones.}

We also apply the above moving sample analysis to the 100 simulated samples (created in section \ref{result.rev_prcor.eff_obsbias}) to investigate the {\xie effects of random pairing and} observational biases. The blue(red) shaded region in \reffig{mov_sample_test} (both panel) shows the $68.3\%$ confidence interval of the results for Kendall(Pearson) test. As expected from \reffig{simu_pr_dis}, most simulated samples have large P values, and thus not likely to produce the observed correlation nor the transition between correlated and uncorrelated. 

In \reffig{diff_subsample_size}, we compare the results of changing the moving sample size from 15 to 16 and 17. {\jiang In this figure, we can see the results are similar, which demonstrates that the result is not sensitive to a specific bin size.}

%小结
As a summary of the moving sample analysis, we find an evidence of orbital spacing dichotomy, namely, orbital period ratios are significantly correlated for tightly packed systems but nearly uncorrelated for loosely packed systems. 
The boundary of such a dichotomy is around $\overline{PR}\sim1.5-1.7$, {\xie i.e, the grey shaded area in \reffig{mov_sample_test}}.
In fact, this dichotomous feature can also be seen from the envelope of the data (see \reffig{diff_pop_prcor}).

\begin{figure}%[tbhp]
\centering
\includegraphics[width=\linewidth,height=0.6\linewidth]{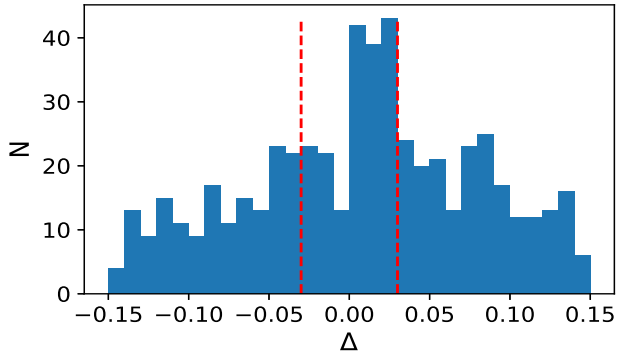}
\caption{{\xie The distribution of $\Delta$ of neighbouring pairs in Kepler multiple transiting systems. We can see the overabundance of planet pairs just outside exact mean motion resonances (MMR) as in \citet{Lis11,Fab14}. We set the boundary of MMR proximity as $|\Delta| < 0.03$ (vertical dashed lines) to include the peak of the overabundance.}}
\label{MMR_criterion}
\end{figure}

\begin{figure*}%[tbhp]
\centering
\includegraphics[width=1\linewidth,height=1.2\linewidth]{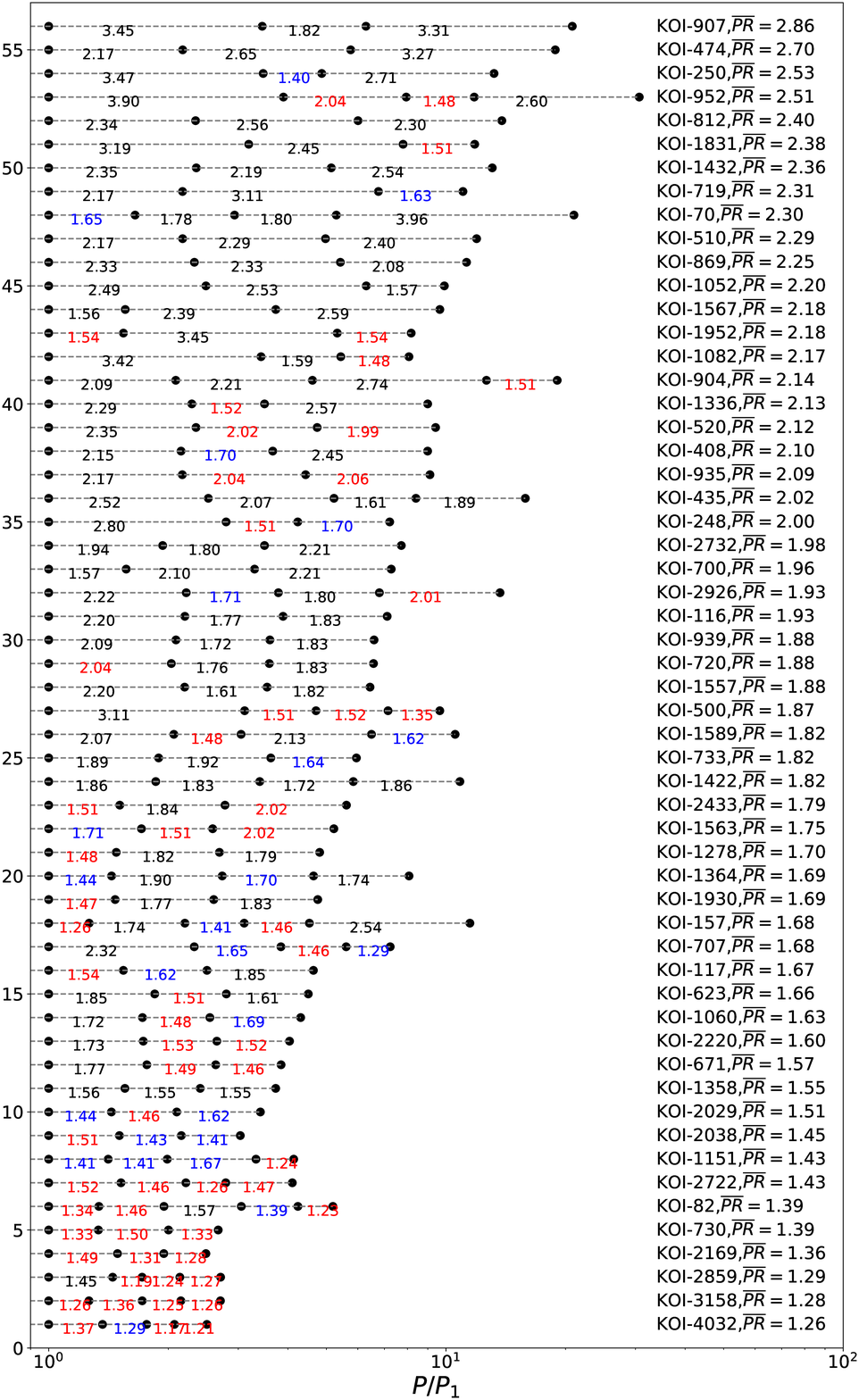}
\caption{Overview of the orbital architectures of planetary systems in the nominal sample.  Each dot denotes a planet or planet candidate and each line of dots represents a planetary system with its name on the right edge of the figure. The orbital periods of the planets are normalized by the orbital periods of the innermost planets in the same systems. Between each pair of adjacent planets, there is a number indicating the orbital period ratio. The red color denotes the proximity to first order mean motion resonances (MMRs) and the blue to second order MMRs.}
\label{sys_overview}
\end{figure*}

\begin{figure}%[tbhp]
\centering
\includegraphics[width=\linewidth,height=\linewidth]{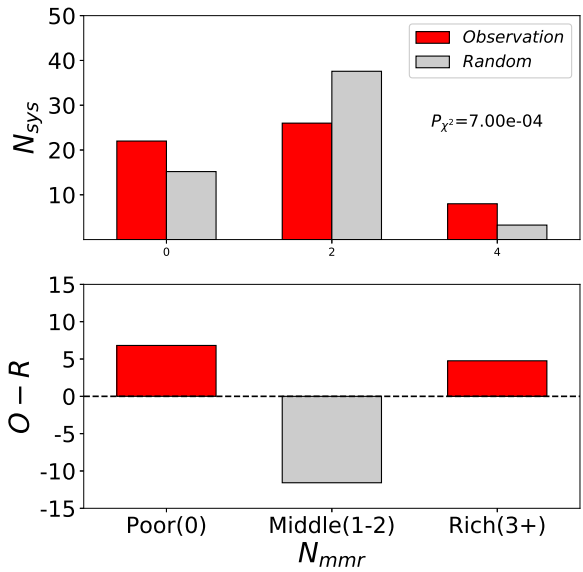}
\caption{The number distributions of MMR poor, middle and rich systems (defined in section \ref{discuss.interp.mmr_dich}) in both the nominal sample (red) and expected values from the corresponding random simulations (grey). The P value of Chi-square test, $P_{\chi^{2}}=7\times10^{-4}$ is printed on the upper panel. In the bottom panel, the difference in $N_{sys}$ between the nominal observed sample and the simulated one are plotted. As can be seen, there is excesses in both MMR poor and rich systems and an deficit in MMR middle system in the observed sample.}
\label{MMR_dich}
\end{figure}

\begin{figure}%[tbhp]
\centering
\includegraphics[width=\linewidth,height=0.9\linewidth]{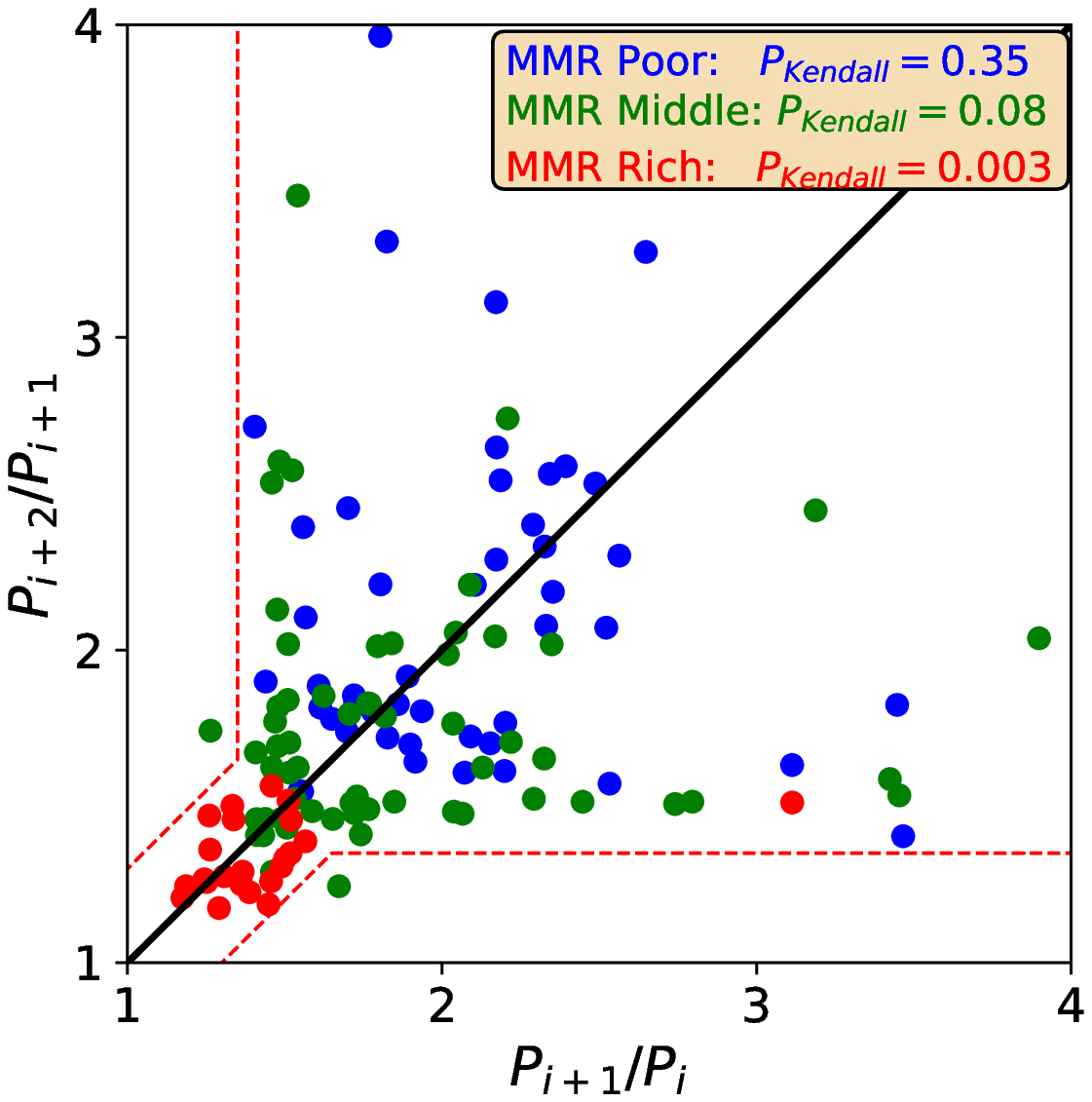}
\caption{{\xie Similar to the upper-left panel in Figure \ref{obs_pr}, but here we divide the nominal sample into three subsamples, MMR poor (blue), middle (green) and rich (red) (see section \ref{discuss.interp.mmr_dich}). 
Fore each subsample, we repeat the Kendall correlation test and print the corresponding P value, $P_{Kendall}$.
As can be seen, the period ratio correlation is significant ($P_{Kendall}=0.003$) in MMR rich systems, but weak in MMR middle ($P_{Kendall}=0.08$) and MMR poor ($P_{Kendall}=0.35$) systems.
The two broken dashed lines generally match the envelopes of the data.
The break points are at $PR=1.65$, which are consistent with the transition zone ($PR=1.5-1.7$) in \reffig{mov_sample_test} (see \ref{discuss.interp.project} for more discussion). 
}}
\label{diff_pop_prcor}
\end{figure}

\section{Discussions} \label{discuss}
%s承前启后
In this work, we have revisited the period ratio correlation of Kepler multiple transiting systems.
    {\xx Unlike the bootstrap method based on the observed systems in \citet{Wei18}, we take a different approach by generating the intrinsic planet populations and forward modeling the transit detection process.
Our forward modeling approach naturally takes into account various effects in the process, such as the effects of transit detection efficiency, orbital stability (as concerned by \citet{Zhu20}) and missing planets (see section \ref{discuss.interp.ntp} below).
}
{\xie
We confirm that the period ratios are, in general, indeed correlated, which cannot be explained by selection effects from observational bias.
Our result is consistent with that of \citet{He19}, which also took a forward modeling approach and  found that the observed distributions of ratios of period ratios are more peaked around unity than their model prediction if assuming no correlation between period ratios at all.
}

Furthermore, we have revealed that the period ratio correlation is highly dependent on period ratio itself, and it shows a dichotomous feature, namely, {\jiang the correlation is strong only in tightly packed systems and becomes weak in loosely packed ones (\reffig{mov_sample_test}).}
In the following, we discuss the implications of such an orbital spacing dichotomy. 
Specifically, we present our interpretation in section \ref{discuss.interp}, and then discuss some future tests to this interpretation in section \ref{discuss.pred}. 
%{\xie Some other effects are discussed in \ref{discuss.other}}. 

\subsection{Interpretation} \label{discuss.interp}
As shown in \reffig{mov_sample_test}, the boundary of the period ratio correlation dichotomy is around period ratio $\sim1.5-1.7$. 
Is this a coincidence?
In the following, we interpret this as a result {\jiang that might be related to } Mean Motion Resonance (MMR) distribution.

\subsubsection{MMR Dichotomy} \label{discuss.interp.mmr_dich}
%%% 介绍Figure 7 和 near MMR criterion
Following \citep{Lit12}, we use the parameter $\Delta$ to describe the proximity of a period ratio to $j+1:j$ MMR,
\begin{equation}
\Delta=\frac{j}{j+1}{\rm PR}-1,
\end{equation}
where $\rm PR$ is the period ratio of adjacent planets.
{\xx Note that $\Delta$ is calculated with respect to the nearest first order MMR. (making the absolute value of $\Delta$ a minimum.)}
{\xie Figure \ref{MMR_criterion} shows the $\Delta$ distribution of neighbouring planet pairs in the Kepler multiple transit systems. Similar to \citet{Lis11} and \citet{Fab14}, we also see an overabundance just outside the MMR center (i.e., $\Delta=0$).
As the overabundance is mainly within $|\Delta|=0.03$, therefore, we set it as the boundary to select those near-MMR period ratios.
}

We plot in \reffig{sys_overview} an overview of the orbital architecture of the planetary systems in our nominal sample.
Each dot denotes a planet or planet candidate, and each line of dots represents a planetary system with its name on the right edge of the figure.
The orbital periods of the planets are normalized by the orbital periods of the innermost planets in the same systems. 
Between each pair of adjacent planets, there is a number indicating the orbital period ratio.
{\jiang All the systems are sorted bottom-up according to the average period ratios, $\overline{PR}$.
We have an intuitive impression that near-MMR period ratios are clustered in compact systems rather than randomly and evenly distributed among all systems.}
In order to see how the distribution of near-MMR period ratios deviates from random distribution, we perform the following statistical test.  
{\xie Note, in the following analysis, we consider only the first order MMRs for simplicity as we found that the result would be similar if the second order MMRs were included.}

We classify all the planetary systems into three groups according to the number of near-MMR pairs in each system: MMR poor (zero near-MMR pair), MMR middle (one or two near-MMR pairs) and MMR rich (three or more near-MMR pairs) systems.
For our nominal sample (sample 1), the numbers of MMR poor, MMR middle and MMR rich systems are 22, 26 and 8 respectively. 
We then apply the same classification to those 10000 randomly simulated systems where near-MMR period ratios are randomly distributed. 
The average number (expectation) is 15.2, 37.6 and 3.2 in MMR poor, MMR middle and MMR rich systems, respectively.
These results are plotted in \reffig{MMR_dich}.
In the top panel, we count the number of systems of these three groups.
We compare the observed numbers (red histogram) with what we would expect (grey histogram) if all near-MMR pairs are randomly distributed.
The chi-square test gives a $\chi^2=13.286$ for the deviation of the observed sample from the expectation, and there are only 7 in 10000 times of random realizations resulting larger $\chi^2$. 
This gives a P value of $7\times 10^{-4}$, indicating  that the distribution of near-MMR period ratios significantly deviates from a random distribution. 
As can be seen from the bottom panel, with respect to random distribution, the distribution of near-MMR period ratios is polarized into the two ends: MMR rich and MMR poor.
There is a deficit in MMR middle class systems. 

Note, MMR is loosely defined here, namely, it generally refers to planets pairs whose period ratios are close to MMR, regardless of whether they are dynamically in MMR state with librating resonant angles. {\xie Previous studies \citep{Lis11,Fab14} have shown that the \emph{global} period ratio distribution deviates somewhat from random distribution in the sense that there is an overabundance of near-MMR ones. Here, we further show that the \emph{local} period ratio distribution also deviates from random distribution, namely, those near-MMR period ratios are not evenly distributed among individual systems. 
Some systems are MMR rich, while some are MMR poor, forming a MMR dichotomy (\reffig{MMR_dich}).}

{\xie
\subsubsection{PR dichotomy or MMR dichotomy ?} \label{discuss.interp.project}
So far, we have revealed two dichotomous features on the orbital spacing, i.e., the period ratio (PR) dichotomy and the MMR dichotomy.
In fact, the two dichotomies are largely equivalent to each other.
On one hand,  MMR dichotomy could be nothing more than a restatement of the PR dichotomy (the small period ratio correlation) given the fact that MMRs are denser for smaller period ratios.

On the other hand, the apparently small period ratio correlation (i.e., PR dichotomy) could also be just a projection of the MMR dichotomy.
%进一步展开论述
As shown in \reffig{sys_overview}, most of the first order and second order MMRs (except for the 2:1 MMR and 3:1 MMR) have period ratios in a relatively small range ($PR\le5:3 \sim 1.7$, ).
Thus, period ratios of a MMR rich system are more likely to be correlated to each other, while such a correlation is not expected in a MMR poor system, causing the apparent PR correlation dichotomy. 
These are clearly shown in \reffig{diff_pop_prcor}.
As can be seen, the two broken dashed lines generally match the envelopes of the data in \reffig{diff_pop_prcor}.
The envelopes of MMR rich systems generally follow the parts that are parallel to the 1:1 line, and thus resulting in strong PR correlation with a P value of Kendall correlation test of $P_{Kendall}=0.003$.
In contrast, the envelopes of other systems generally follow the part that are parallel to the x and y axes respectively, resulting in weak PR correlations in MMR middle ($P_{Kendall}=0.08$) and MMR poor ($P_{Kendall}=0.35$) systems.
The break points of the dashed lines are at $PR=1.65$, which are consistent with the transition zone ($PR=1.5-1.7$) as shown in \reffig{mov_sample_test}.
}

{\xie That being so, then which one is more essential to reflect the orbital spacing pattern? PR dichotomy or MMR dichotomy? Here, we prefer the MMR dichotomy rather than the PR dichotomy for the following reasons.}

{\xie First, PR dichotomy or small period ratio correlation is just a mathematical correlation whose boundary ($\overline{PR}\sim 1.5-1.7$, \reffig{mov_sample_test}) itself needs an additional explanation, while the MMR dichotomy is more physically-based and naturally explains the PR correlation boundary (as discussed above and shown in \reffig{diff_pop_prcor}}).

{\xie Second, perhaps more importantly,} the MMR dichotomy could be a natural result of planet migration and dynamical evolution. 
%进一步展开论述 
One of the leading models on the formation of close-in super-Earths is the inward migration model, namely planets formed at larger distances (e.g., snowline) from the star followed by inward migration driven by gas disk \citep{TP07,IL08,Cos14,HN12}.
At the beginning when the gas disk was present, planets grew and migrated inward to form a MMR chain.
Afterwards, when the gas disk dissipated, these MMR chains generally evolved to the following two branches \citep{Izi17}.
On one hand, some of the MMR chains could become dynamically unstable, which underwent a phase of giant impact that erased the footprint of MMR.
On the other hand, some MMR chains could remain relatively stable. 
Although most of these MMRs could still be broken afterwards due to various mechanisms e.g., tides damping \citep{LW12,BM13,DL14}, planetesimal interaction \citep{CF15} and etc., many of these effects are gentle and planets are able to stay near MMR with approximately commensurable period ratios. 
These two branches of dynamical evolution naturally lead to the MMR dichotomy

{\xie As a conclusion of above discussions, we therefore consider the orbital spacing pattern dichotomy shown in \reffig{mov_sample_test} is a consequence of the MMR dichotomy (\reffig{MMR_dich}.}

\begin{figure*}%[tbhp]
\centering
\includegraphics[width=1\linewidth,height=0.32\linewidth]{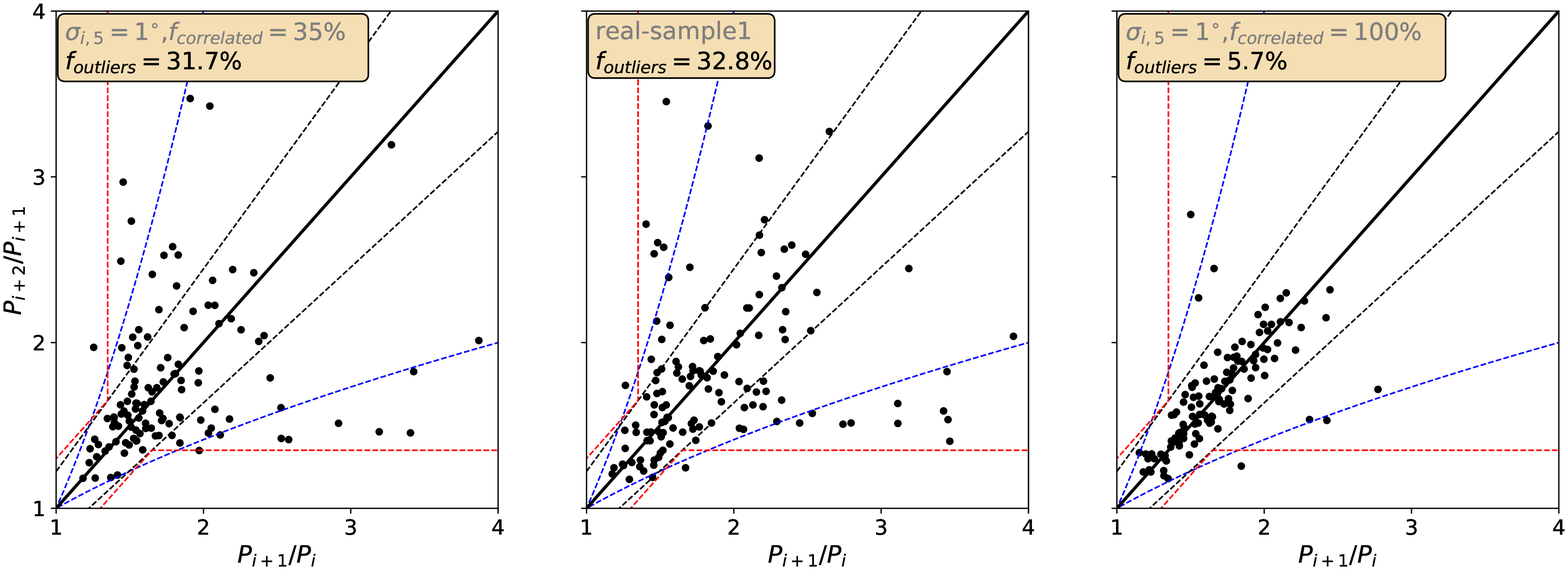}
\caption{{\xx Ratio distributions of Period ratios for the one-population scenario (right panel, all planets are generated from a period ratio correlated population, i.e., $f_{correlated}=100\%$), twp-population scenario ($f_{correlated}=35\%$, left panel) and the real sample 1 (middle panel, the same data as Figure \ref{diff_pop_prcor}).  
In each panel, The black solid line shows perfect correlation, i.e., $y=x$. 
The two black dashed lines $y=\frac{11}{9}x$ and $y=\frac{9}{11}x$ represent 10\% deviation from perfect period ratio correlation).
The two blue dashed lines $y=x^2$ and $y=\sqrt{x}$ denote the expected locations of outliers caused by missing the intermediate planets. 
In each panel, we print the fraction ($f_{outliers}$) of outliers, namely the data points further away from the perfect correlation line, i.e.,y=x, than the two black dashed lines.
See more details in section \ref{discuss.interp.ntp}.
}}
\label{twopop_dis}
\end{figure*}

\begin{figure}%[tbhp]
\centering
\includegraphics[width=\linewidth,height=0.8\linewidth]{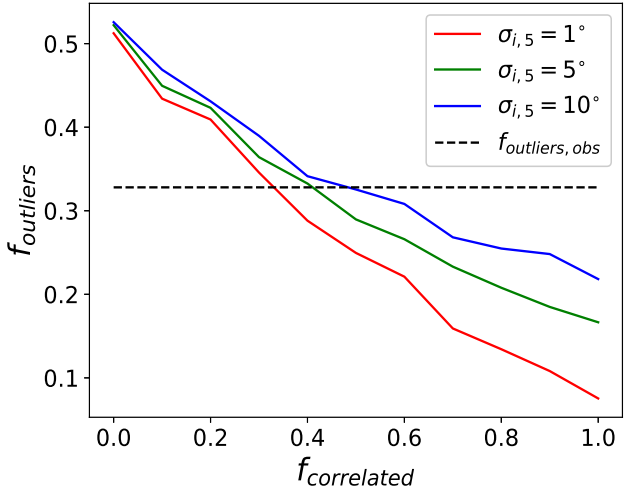}
\caption{{\xx The fraction of outliers $f_{outliers}$ as a function of the fraction of the correlated population $f_{correlated}$ in a two-population scenario, with different inclination dispersion $\sigma_{i,5}$ parameters. 
To reproduce a similar $f_{outliers}=32.8\%$ as in the observed sample 1 (black-dashed line), the $f_{correlated}$ should be around $\sim$35\% (for $\sigma_{i,5}=1^{\circ}$) to $\sim$50\% (for $\sigma_{i,5}=10^{\circ}$). 
}}
\label{twopop_trend}
\end{figure}

{\xx \subsubsection{Effect of Missing Planets \label{discuss.interp.ntp}}
Planets which intrinsically exist between the detected transiting planets could be missed by the transit survey, due to either weak signals (low SNR) or non-transiting geometry.
In our forward modelling simulations, we found $\sim$2-3\% of planets in the simulated transiting multi-planet systems were missed due to low SNR, and $\sim$ 1\%-14\% (depending on the intrinsic inclination dispersion, $\sigma_{i,5}$) of them were missed because of non-transiting geometry .
These missing planets cause the observed period ratios larger than the intrinsic ones, which randomizes the period ratio distribution to some degree.
If adopting a typical minimum intrinsic period ratio of $1.2$, this effect can affect period ratios larger than $1.2^2\sim$ 1.4. 
Therefore, one might concern that the observed tendency of weaker period ratio correlation at larger period ratios could be caused by the effect of missing planets.
In the follows, we quantify this effect.

First, we investigate a one-population scenario with a toy model, in which period ratios are intrinsically correlated (along the diagonal line of Figure \ref{diff_pop_prcor}), and those observed uncorrelated period ratios (outliers away from the diagonal line) are caused by the missing planets.
Specifically, to generate an intrinsic system, we randomly draw the first period ratio from the debiased period ratio distribution (Appendix 1), then draw other period ratios with a random deviation within 10\% from the first one. 
A typical result of the one-population scenario is shown in the right panel of Figure \ref{twopop_dis} with  $\sigma_{i,5}=1^{\circ}$ and $f_{correlated}=100\%$ (i.e., 100\% systems are period-ratio-correlated).
As compared to the result of real sample shown in the middle panel, the one-population scenario fails to reproduce the observation in the following two aspects. 
\begin{enumerate}
    \item It produces too few outliers (points away from the diagonal line further than the two dashed lines, ($11y=9x$ and $9y=11x$ in a x-y plane, see the caption of Figure \ref{twopop_dis}).
    The outliers fraction is 5.7\% vs. the observed 32.8\% in this case.
    Although increasing the intrinsic orbital inclination dispersion $\sigma_{i,5}$ generally increases the numbers of non-transiting planets and thus the fraction of outliers, it is still significantly lower than the observed one even if assuming an unrealistically large $\sigma_{i,5}=10^\circ$ (as shown in the bottom right part of Figure \ref{twopop_trend}).
    
    \item Its envelopes (set by the outliers), as expected, follow the blue dashed lines in Figure \ref{twopop_dis} ($y=x^2$ and $y=x^{0.5}$ in a x-y plane), which is significantly different from the observed one (red dashed lines in Figures \ref{diff_pop_prcor} and \ref{twopop_dis}).
\end{enumerate}

Second, we then further consider a two-populations scenario with a toy model, in which only a fraction ($f_{correlated}<100\%$) of systems are assumed as period-ratio correlated as in the above one-population scenario. 
For the other $1-f_{correlated}$ fraction of systems, the period ratios are randomly drawn from the debiased period ration distribution but with a lower limit truncated at 1.35 (motivated by the apparent envelopes). 
As shown in Figure \ref{twopop_trend}, by adding more uncorrelated population systems (i.e., decreasing $f_{correlated}$), the outlier fraction generally increases, and it meets the observed value if $f_{correlated}\sim35\%$ for $\sigma_{i,5}=1^\circ$.
In the left panel of Figure \ref{twopop_dis}, we plot the ratio distribution of  period ratios for this specific case.
As can be seen, the two-populations toy model largely reproduces the result of the real observed sample, especially in terms of both the outlier fraction (31.7\% vs 32.8\%) and the distribution envelopes. 

As a summary of this subsection, we conclude that the effect of missing planets (either low SNR planets or non-transiting planets) alone is too small to reproduce the observed ratio distribution of period ratios (Figure \ref{twopop_dis}).
In addition, we find that the observed results could be largely reproduced with a two-populations toy model, which further demonstrates the dichotomy nature of the orbital spacing pattern.
}

\subsubsection{Effect of Ultra Short Period Planets} \label{discuss.interp.usp}
Systems with ultra short period (USP, period $<1$ day) are found to have relative larger period ratios \citep{WF15} and larger orbital inclinations \citep{Dai18}, and they could have undergone some different formation history \citep{Pet19,PL19}.
Thus, one might concern whether USP planets are related to the observed trend of weaker period ratio correlation in systems with larger period ratios.
However,the occurrence rate of USP planets is in fact very low ($\sim0.5\%$) around sun-like stars \citep{San14}. 
In our nominal sample, only 2 out of 56 systems host USP planets.  
After removing these two systems, we repeat the moving sample analysis and find that the result is nearly unchanged as compared to Figure 4. 
We therefore conclude that our results are not affected by USP planets.

\subsection{Predictions} \label{discuss.pred}
%承上启下
Based on the above discussions on the dynamical origin of the MMR dichotomy, we may further make some predictions for future studies.

First, we predict that the planets in  MMR-poor systems (with relatively larger and thus uncorrelated period ratios) may have larger masses, densities and orbital eccentricities/inclinations than those in MMR-rich systems (with relatively smaller and thus correlated period ratios).
This is simply because the giant impact process which erased the footprint of MMR also increased the masses and the orbital eccentricities/inclinations of planets.
The prediction on mass and density is consistent with the recent finding that the masses and densities of TTV ({\jiang Transit Timing Variation}) planets (most are near MMR) are systematically lower than those of the RV (radial velocity) planets (most are not near MMR) \cite{Ste16}.
The confirmation of the prediction on orbital eccentricity/inclination is not trivial, because the increase in eccentricity/inclination is moderate, which requires future dedicated studies on orbital characterization.

Second, we may predict that MMR-poor systems (with relatively larger and thus uncorrelated period ratios) are relatively older than those MMR-rich systems (with relatively smaller and thus correlated period ratios). 
This is simply based on the consideration that the longer time of dynamical evolution (e.g., giant impact, tidal damping and planet-planetesimal interaction), the larger probability to erase the footprint of MMR.
The prediction on age is qualitatively consistent with the result of previous study \citep{KZ11} based on the radial velocity planet sample.
Future studies with large and diverse samples are needed to fully establish this point.

\section{Summary} \label{summary}
In this paper, we studied the pattern of orbital spacings (in terms of period ratios) of Kepler multiple planet systems. 
We confirm that, period ratios are indeed somewhat correlated (Figure \ref{obs_pr}), and such a correlation is unlikely to be caused by observational biases (Figures \ref{typi_simu_pr}-\ref{simu_pr_dis}). 
Furthermore, we reveal that the above orbital spacing pattern is dichotomous, namely, {\xie period ratios are strongly correlated to each other in the tightly packed systems, but uncorrelated at all in the loosely packed systems. 
The transition from correlation to noncorrelation is abrupt with the boundary at  $ Median(\overline{PR}) \sim 1.5-1.7$ (section \ref{result.evi_prdich} and Figure \ref{mov_sample_test}).}

Then, we relate such a period ratio dichotomy to another dichotomy that reflects the near-MMR period ratios tend to be clustered rather than evenly distributed (dubbed as MMR dichotomy for short, see section \ref{discuss.interp} and Figures \ref{sys_overview}-\ref{MMR_dich}). 
The MMR dichotomy naturally leads to a transition from period ratio correlation to non-correlation around $\overline{PR}\sim1.5-1.7$ (\reffig{diff_pop_prcor}), and it could be also a natural result of planet migration and dynamical evolution (section \ref{discuss.interp.project}).
{\xx The transition from period ratio correlation to non-correlation cannot be explained by the missing intermediate planets (due to either low SNR or non-transiting geometry,  section \ref{discuss.interp.ntp}) nor by ultra short period planets (section \ref{discuss.interp.usp}).
Nevertheless, it can be largely reproduced with a two-population toy model, further demonstrating the dichotomy nature of the orbital spacing pattern.} 

Finally, based on the formation of the MMR dichotomy, we predict that planets in MMR-poor systems are more massive, denser and dynamically hotter (larger orbital eccentricities and inclinations) than those in MMR-rich ones (section \ref{discuss.pred}). 
%In addition, we discussed some other effects , e.g., non-transiting planets and ultra short period planets, and conclude that they have little effect on our results (section \ref{discuss.}).

\label{sec:acknowledgments}
\acknowledgments
We thank W. Zhu for helpful comments and suggestions. This work is supported by the National Key R\& D Program of China (No. 2019YFA0405100) and the National Natural Science Foundation of China (NSFC) (grant No. 11933001). J.-W.X. also acknowledges the support from the National Youth Talent Support Program and the Distinguish Youth Foundation of Jiangsu Scientific Committee (BK20190005)

\clearpage
\appendix
\section{Monte Carlo simulations of Transit Systems}
\label{MC_trasys}
In order to quantify the probability of reproducing the observed period ratio correlation by observational bias, we perform Monte Carlo simulations of transit systems by the following forward modeling of transit observations.
Specifically, first (section \ref{MC_trasys.intri_sys}), we generate intrinsic planetary systems based on some reasonable assumptions that are studied and justified by previous studies.
Then (section \ref{MC_trasys.simuobs}), we apply some criteria to simulate transit detection from the above generated systems  .
Finally (section \ref{MC_trasys.evaluate}), we evaluate the period ratio correlation as in section \ref{result.rev_prcor.eval} for the simulated transit sample.
By repeating the above simulation and evaluation 1000 times, we access the probability of reproducing the observed period ratio correlation by observational bias (Figure \ref{simu_pr_dis}).

\subsection{Generating Intrinsic Planet Systems}
\label{MC_trasys.intri_sys}
In the following, we describe the procedure to generate an intrinsic planet system.

\begin{itemize}
%Step 1: Multiplicity

    \item[1.] We randomly select a star from the Kepler input catalog, whose stellar properties have been revised by GAIA data \citep{Ber18}）.

    \item[2.] We assign $K$ planets to the star, where $K =$ 1 - 6 is drawn from the multiplicity function obtained by \citet{Zhu18} (their figure 8).
    
    \item[3.] We draw the orbital period of the innermost planet  randomly from the distribution of orbital periods of innermost transiting planets in the observed sample after correcting the transit geometric bias. To determine the period of other planets in the system, we multiply the period of the inner planet by a period ratio, which is randomly drawn from a distribution debiased from observation using the CORBITS algorithm \citep{BR16, Wu19}. {\jiang Specifically, we calculate the probability of detecting outer planet given that the inner planet is detected. The inverse of the probability is adopted as the weight of the period ratio of the planet pair. }
    
    \item[4.] The radius of each planet is drawn from a debiased radius distribution that is constructed as in \citep{FM12}.
    Specifically, for a planet with radius $R$ and period $P$ in the observed sample, we calculate $\eta$ as the fraction of stars that can detect the transit of such a planet. Since $\eta$ is the ratio of the number of detectable events to the number of actual planets, the inverse of $\eta$ is an estimate of the actual number of planets represented by each detection. 
    Therefore, we set $\frac{1}{\eta}$ as the weight of each observed specific radius $R$ to obtain the debiased radial distribution.
 
    \item[5.] To avoid the cases where two planets are too close to each other and become dynamically unstable, we also adopt the stability criterion as in \citep{FM12} i.e.
    \begin{equation}
    \Delta=\frac{a_2-a_1}{R_{H1,2}}\geq 3.46
    \end{equation}
    where $a_1$ and $a_2$ are the semi-major axis of the inner and outer planet respectively and $R_{H1,2}$ is their mutual Hill radius,
    \begin{equation}
    R_{H1,2}=(\frac{M_{1}+M_2}{3M_\ast})^{1/3}\frac{a_2+a_1}{2}
    \end{equation}
    with $M_1$ and $M_2$ being the mass of the inner and outer planet and $M_\ast$ being the mass of the host star. 
    Masses of planets are estimated using a nominal mass-radius relation \citep{Lis12} i.e.
    \begin{equation}
    \frac{M}{M_\oplus}=(\frac{R}{R_\oplus})^{2.06}
    \end{equation}
    where $M$ and $R$ are the mass and radius of the planet, and $M_\oplus$ and $R_\oplus$ is the mass and radius of Earth, respectively.
    
    \item[6.] For each system that passed the orbital stability check, we assign $I_p$, the orbital inclination relative to the observer to the planets. Following \citet{Zhu18}, in practice,  we calculate
    \begin{equation}
    \cos I_p = \cos I \cos i \ -\ \sin I \sin i \cos \phi,
    \end{equation}
    where $I$ is the inclination of the system invariable plane, $i$ the planet inclination with respect to this invariable planet, and $\phi$ the phase angle. The distribution of $I$ is isotropic (i.e., $\cos I$ is uniform for $0^\circ <I<180^\circ$ ) and $\phi$ is a random between 0$^\circ$ and 360$^\circ$. For single planet systems, $i=0^\circ$ and $I_p=I$. For multiple planet systems, following \citet{Zhu18}, $i$ is modeled as a Fisher distribution,
    \begin{equation}
    P\left(i|\kappa_k\right)=\frac{\kappa_k \sin i}{2 \sinh \kappa_k}e^{\kappa_k \cos i}.
    \end{equation}
    The $\kappa_k$ parameter is related to the inclination dispersion as
    \begin{equation}
    \sigma_{i,k}^2 = \left<\sin^2 i\right> = \frac{2}{\kappa_k}\left(\coth \kappa_k - \frac{1}{\kappa_k}\right).
    \end{equation}
    Here, also following \citet{Zhu18}, the inclination dispersion is a power law function of the planet multiplicity, $k$, 
    \begin{equation}\label{eq:inclination}
    \sigma_{i,k}\equiv \sqrt{\left<\sin^2i\right>}=\sigma_{i,5}\left(\frac{k}{5}\right)^{\alpha}.
    \end{equation}
    Here, we adopt the typical results from \citet{Zhu18}, i.e., $\sigma_{i,5}=0.8^\circ$ and $\alpha=-4$ . 

\end{itemize}

\subsection{Simulating Transit Observation}
\label{MC_trasys.simuobs}
We first consider the transit geometric effect.
A transit is defined as the impact parameter less than 1, i.e., $|\cos \left(I_p\right)/\epsilon|<1$, where $\epsilon=R/a$ is the transit parameter. As in \citet{Zhu18}, we ignore the minor impact of the planet size and eccentricity.

We than consider the effect of detection efficiency.
Specifically, we remove the non-detectable transiting-planets with transit Signal to Noise Ratio (SNR) lower than 7.1 according to \citet{Mul15}.
Following \citet{Nar18}, the transit SNR is calculated as
\begin{equation}
    SNR=(\frac{R}{R_\ast})^2\frac{\sqrt{N}}{\sigma_{CDPP}}
\end{equation}
where $R$ and $R_\ast$ are the radii of planet and star respectively and $N$ is the effective transiting times.
$\sigma_{CDPP}$ represents the combined differential photometric precision of the star.

\subsection{Evaluation of PR Correlation}
\label{MC_trasys.evaluate}
We repeat above procedure until obtaining the same number of simulated transiting systems after the same filters as the observed ones (table \ref{Tab.samp_des} in section \ref{sample}). 
For the four simulated samples, we perform the same period ratio correlation evaluation as for the observed ones (section \ref{result.rev_prcor}).
The typical results are illustrated in Figure \ref{typi_simu_pr}.
As can be seen, the P value, $P_{Kenall}$ and $P_{Pearson}$, of all the four simulated samples are of the magnitude of $10^{-1}$, which are consistent with no period ratio correlation.
The $P_{Kendall}$($P_{Pearson}$) distributions of 1000 Monte Carlo realizations are plotted in Figure \ref{simu_pr_dis}.
For samples 1,2 and 4, the simulations lead to  $P_{Kendall}$ ($P_{Pearson}$) larger than that of the observed one in most cases. 
Therefore, the period ratio correlations observed in the samples 1,2 and 4 are likely to be physical rather than the results of observational biases.

Note, although our model is relatively simple and suffer some uncertainties, for example, the intrinsic multiplicity is actually not well constrained and the transit detection efficiency is considered as a simple SNR cut, it catches the bases of transit simulation.
A more sophisticated state of art model may improve the estimate the planet occurrence rates, but it is unlikely to change the conclusion, namely, the process that generates transit systems cannot produce significant period ratio correlation.

\bibliography{main}

\end{document}